\documentclass{mem}
\usepackage{natbib}\usepackage{txfonts}\usepackage{balance}
\usepackage{graphicx}
\idline{75}{282}
\begin{document}

\title{Galactic Nuclear Cluster Formation Via Globular Cluster Mergers}

   \subtitle{}

\author{R. \,Capuzzo--Dolcetta\inst{}}

  \offprints{R. Capuzzo--Dolcetta}

\institute{Dipartimento di Fisica --
Sapienza, Universit\`a di Roma, piazzale A. Moro 2,
I-00185 Roma, Italy
\email{roberto.capuzzodolcetta@uniroma1.it}
}

\authorrunning{Capuzzo--Dolcetta}

\titlerunning{Nuclear cluster formation}

\abstract{We apply the idea that dense stellar systems in the central region of galaxies are formed via globular cluster mergers to the formation of the nuclear star cluster of the Milky Way, where a massive black hole (Sgr A$^*$) is present.
Our high precision $N$- body simulations show a good fit to the observational characteristics of the Milky Way nuclear cluster, giving further reliability to the so called migratory model for the formation of compact systems in the inner galaxy regions

}
\maketitle{}

\section{Introduction}
 
Compact Massive Objects (CMOs) are almost ubiquitously present in the galactic centers, in form of massive or supermassive black holes (MBHs, SMBHs) or in a more {\it dilute} form, i.e. resolved stellar nuclei or Nuclear Star Clusters (NSCs).
There is a direct correlation between compactness of the CMO and the host galaxy luminosity: the brighter the host the denser the CMO, till the limit of BH infinite density (see, e.g., \citet{vol08}). 

An interpretation of the modes of formation of CMOs in galaxies is surely a  modern topic, which have not yet received an adequate self-consistent explanation frame.
Preliminarily, we must say that a common origin for the CMOs in galaxies, with a second order difference in their compactness and mass as due to properties of the host galaxy, is an appealing view, but faces with the evidence, for example, that some galaxies contain both an NSC and a SMBH \citep{boe08}. 
A well-known example is the Milky Way (MW), where a $4\times 10^6$ M$_\odot$ black hole coexists with an NSC $\sim 4$ times more massive (M$_{NSC} \approx 1.5\times 10^7$ M$_\odot$); the same coexistence has been found recently in various other cases.
Anyway, in spite of these peculiarities, it is surely intriguing the possibility, raised by \citet{cap93} and furtherly developed by him and collaborators, that CMOs have formed by stellar aggregations like massive globular clusters (GCs) brought to the host galaxy center via dynamical braking (dynamical friction). The infalling and orbitally decaying GCs eventually merge in the innermost region of the host galaxy where they form an actual Super Star Cluster in times and modes that are, partly, governed by the galaxy structure, i.e. geometrical shape, steepness of mass density distribution as well as total mass.
If the two phenomena of dynamical friction braking, yielding to a central mass accumulation, and tidal disruption, working in the opposite direction, are the leading mechanisms, the host galaxy luminosity (mass) correlation with the CMO compactness has a qualitative explanation. 
Actually, fainter galaxies show a higher central density in phase space, thus enhancing the effect of dynamical friction on GCs and favoring their accumulation to grow a NSC, while bright and very bright galaxies borrow, usually, MBHs and SMBHs in their center which may act as destroyers of incoming GCs inibiting the local NSC formation.

\begin{figure}[t!]
\resizebox{\hsize}{!}{\includegraphics[clip=true]{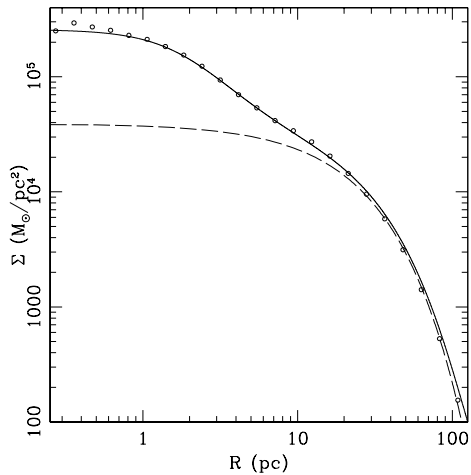}}
\caption{\footnotesize
Projected density profile at the end of the merger simulation. The solid line gives the best-fitting model to the entire system (galaxy+NSC), and the dashed curve gives the fit to the density profile of the galaxy (from Aetal12). 
}
\label{sdens}
\end{figure}

\begin{figure}[t!]
\resizebox{\hsize}{!}
{\includegraphics[clip=true]{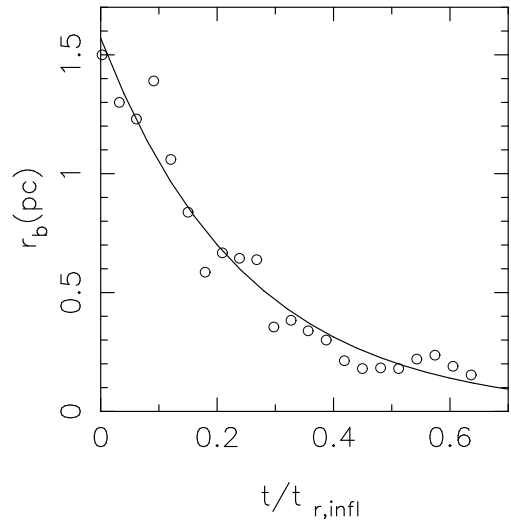}}
\caption{\footnotesize
Post-infall evolution of the $N$--body model. The plot shows the break (core) radius as a function of time (in units of the relaxation time at the BH influence radius). The solid line is the best-fit exponential (from Aetal12).
}
\label{pinfall}
\end{figure}

\section{The formation of the MW Nuclear Star Cluster}
The possibility of cumulation of a significant amount of mass in the innermost regions of galaxies in form of infalling GCs is an old idea by \citet{tre75}.
For some time, the dynamical friction effect was considered important just for very massive GCs, until its role, enhanced in triaxial galaxies \citep{pes92} and in cuspy galaxies \citep{vic05,arc12}, was clearly shown to be relevant for a wide range of GC structural parameters and initial conditions by \citet{cap93} and later papers.
This framework of GC infalls provides possible answers on both the origin of local black hole mass feeding and on the modes of formation of the observed dense stellar systems in both dwarf ellipticals and spirals. Actually, these systems have characteristics similar to those of GCs but on a magnified scale. This {\it migratory} explanation is alternative to that claiming a funneling of molecular gas toward the inner part of galaxies and subsequent local star formation (see for instance \citet{loo82}). 
The two hypotheses have been differently tested by various authors; the {\it dissipationless} hypothesis (GC migratory model) has also been quantitatively well tested by specific $N$--body simulations while the {\it  dissipative} (gaseous) origin has not yet been thoroughly investigated.

It seems, so, logical performing a test of the migratory model for the origin of galactic central dense stellar systems in the best known environment: the Milky Way (MW), where, around the $\sim 4\times 10^6$ M$_\odot$ central black hole there is a NSC of about $10^7$ M$_\odot$ whose photometric structure as well as kinematical characteristics are studied at a level unreachable in external galaxies. As a matter of fact, the MW NSC is, so far, the only one resolved in its stellar content in spite of its high density.
As a consequence, in a recent paper \citet{ant12} (hereafter Aetal12) tackled the problem of analyzing whether the migratory hypothesis for the formation is working also in the specific context of the MW, where, indeed, the dynamical role of the black hole is likely to be significant. 

The t observational constraints are: 
\par\noindent
$i)$ the mass of the NSC ($\sim 1.5\times 10^7$ M$_\odot$);
$ii)$ its mass density profile, $\rho(r)$, showing a core in the distribution of the late type stars of $0.5$ pc size, and decreasing further out as $\rho \propto r^{-1.8}$ up to $r\simeq 30$ pc. Out of $30$ pc there is a large nuclear stellar and molecular disk of about same radius;
$iii)$ some kinematic data;
$iv)$ the NSC luminosity function.

We showed how the orbital decay of 12 tidally truncated massive ($\sim 10^6$ M$_\odot$) GCs actually leads, in a sufficiently short time, to the growth and stabilization of a dense stellar cluster around the Sgr A$^*$ BH. We followed the infall and merger of the GCs in two sets of initial conditions (simultaneous and consecutive infalls) and were able to study the density profile shape as well as the distribution in velocity space. Moreover, we studied also the post merger evolution, extending the $N$--body simulations after the final merger for a time that corresponds to about 10 Gyr, after scaling to the MW. The absence of a clear formation of a Bahcall-Wolf cusp in the NSC is not surprising because the expected time scale for its growth is the 2-body relaxation time at the black hole influence radius is in the range $20-30$ Gyr \citep{mer10}.

\begin{figure*}[t!]
\resizebox{\hsize}{!}{\includegraphics[clip=true]{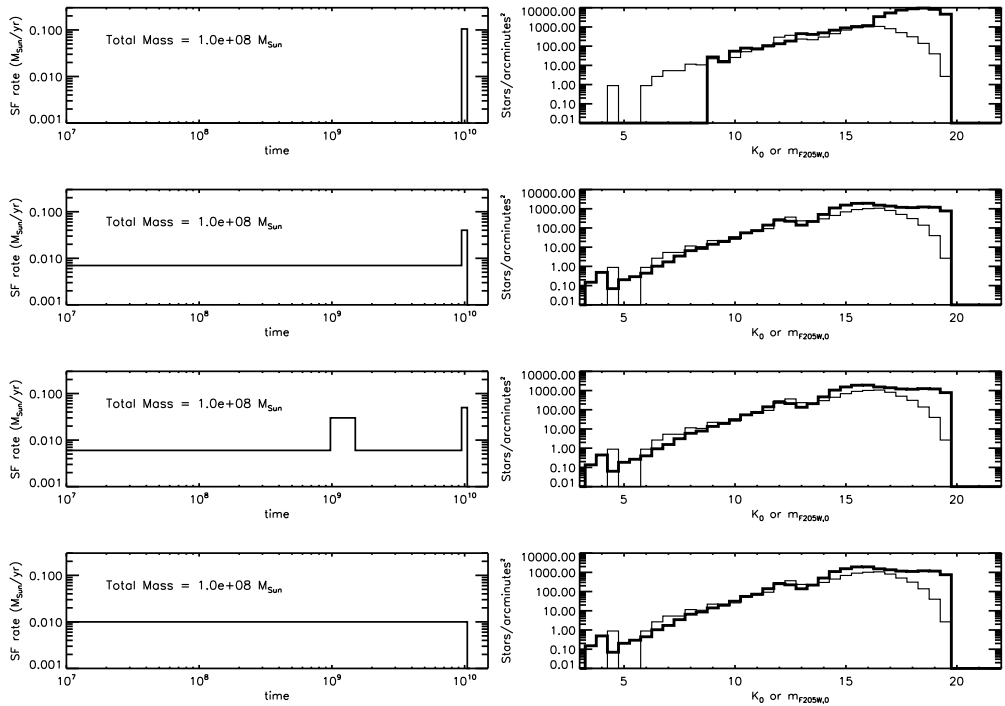}}
\caption{\footnotesize
Left panels display star formation scenarios in which, from top to bottom: the entire mass is built up by ancient stars, brought into the Galactic center by infalling GCs; half of the mass is contributed by old stars and half due to continuous star formation; at the mass contributed by continuous star formation and GCs we add stars formed during a starburst episode occurred at $\sim 1$ Gyr; all the mass is contributed by continuous star formation. Right panels compare the observed LF of the Galactic center (light lines) with the LFs (heavy lines) resulting from the different star formation scenarios assuming solar metallicity and canonical mass-loss rates. Note that the data are much more than 50\% incomplete for the faintest few bins. The models have not been scaled for mass, but rather have been rescaled along the vertical axis to match the number counts in the K = 11.0 bin (from Aetal12). 
}
\label{lfunc}
\end{figure*}

\section{Results}

Taking into account the constraints imposed by the observations of the MW NSC, our results are conveniently summarized as follows:
\begin{itemize}
\item The stellar system resulting from the consecutive mergers
has a density that falls off as $r^{-2}$ and a core of radius
$\sim 1$ pc. These properties are similar to those observed in the
MW NSC (see Fig. \ref{sdens}).
\item The morphology of the NSC evolved during the infalls, from a strongly triaxial shape toward a more oblate/axisymmetric shape. Kinematically, the final system is characterized by a mild tangential anisotropy within the inner 30 pc and a low degree of rotation.
\item The effect of gravitational encounters on the evolution of the NSC was investigated continuing $N$--body integrations after the last inspiral event. The NSC core shrinks by roughly a factor of two in 10 Gyr as the stellar density starts evolving toward a Bahcall-Wolf cusp (see Fig. \ref{pinfall}). This final core size
is essentially identical to the size of the core observed at the center of the Milky Way. The density profile outside the core remained nearly unchanged during this evolution. Gravitational encounters also make the NSC to evolve
toward spherical symmetry in configuration and velocity space.
\item Since GCs are ancient objects with ages $∼10$–$13$ Gyr, in the merger model a large fraction of the NSC mass is expected to be in old stars. Using stellar population models, we showed that the observed LF of the MW NSC is consistent with a star formation history in which a large fraction (about 1/2) of the mass consists of old (∼10 Gyr) stars and the remainder is from continuous star formation (see Fig. \ref{lfunc}).
\end{itemize}

\section{Conclusions}
The migratory-merger model for the formation of dense stellar systems in the centers of elliptical galaxies and of superdense Nuclear Star Clusters in spirals has been applied to the case of the Milky Way Nuclear Star Cluster, where a massive black hole dominates the innermost potential. Our $N$-body simulations show that a NSC with the same spatial and kinematical characteristics of the MW NSC is formed by inspiralling of a limited number ($\sim 10-15$) massive globular clusters in a time short respect to the age of the galaxy. Moreover, the observed luminosity function of the MW NSC is compatible with a younger population emebedded in that coming from the merger object.

\bibliographystyle{aa}

\end{document}